\documentstyle[epsfig,longtable]{aipproc}

\newcommand{\AmS}{{\protect\the\textfont2
  A\kern-.1667em\lower.5ex\hbox{M}\kern-.125emS}}
\newcommand{\fB}{$f_B$}
\newcommand{\fBs}{$f_{B_s}$}
\newcommand{\fD}{$f_D$}
\newcommand{\fDs}{$f_{D_s}$}

\def\3he{{$^3${\rm He}}}

\def\ie{{\it i.e.,\ }}

\def\etc{{\it etc.}}

\def\slD{\raise.15ex\hbox{$/$}\kern-.53em\hbox{$D$}}
\def\dsl{\raise.15ex\hbox{$/$}\kern-.57em\hbox{$\Delta$}}
\def\slp{{\raise.15ex\hbox{$/$}\kern-.57em\hbox{$\partial$}}}
\def\nsl{\raise.15ex\hbox{$/$}\kern-.57em\hbox{$\nabla$}}
\def\sla{\raise.15ex\hbox{$/$}\kern-.57em\hbox{$\rightarrow$}}
\def\slla{\raise.15ex\hbox{$/$}\kern-.57em\hbox{$\lambda$}}
\def\slb{\raise.15ex\hbox{$/$}\kern-.57em\hbox{$b$}}
\def\lnp{\raise.15ex\hbox{$/$}\kern-.57em\hbox{$p$}}
\def\lnk{\raise.15ex\hbox{$/$}\kern-.57em\hbox{$k$}}
\def\lnK{\raise.15ex\hbox{$/$}\kern-.57em\hbox{$K$}}
\def\lnq{\raise.15ex\hbox{$/$}\kern-.57em\hbox{$q$}}

\def\cM{{\cal M}}

\def\pmb#1{\setbox0=\hbox{$#1$}%
\kern-.025em\copy0\kern-\wd0
\kern.05em\copy0\kern-\wd0
\kern-.025em\raise.0433em\box0 }

\def\q2{{Q^2}}
\def\gtwid{\raise.3ex\hbox{$>$\kern-.75em\lower1ex\hbox{$\sim$}}}
\def\ltwid{\raise.3ex\hbox{$<$\kern-.75em\lower1ex\hbox{$\sim$}}}
\def\12{{1\over2}}
\def\part{\partial}

\def\low#1{\lower.5ex\hbox{${}_#1$}}

\def\psl{\raise.15ex\hbox{$/$}\kern-.57em\hbox{$\partial$}}
\def\partt{\raise.15ex\hbox{$\widetilde$}{\kern-.37em\hbox{$\partial$}}}

\def\topppageno1{\global\footline={\hfil}\global\headline
={\ifnum\pageno<\firstpageno{\hfil}\else{\hss\twelverm --\ \folio
\ --\hss}\fi}}
 
\def\toppageno2{\global\footline={\hfil}\global\headline
={\ifnum\pageno<\firstpageno{\hfil}\else{\rightline{\hfill\hfill
\twelverm \ \folio
\ \hss}}\fi}}

\def\prd#1{Phys.\ Rev.\ {\bf D#1}}

\def\ie{{\it i.e.},\ }

\def\nsection#1 #2{\leftline{\rlap{#1}\indent\relax #2}}

\def\prd#1{Phys.\ Rev.\ {\bf D#1}}

\begin{document}
\title{$B$ Mixing on the Lattice: \fB, \fBs\  and Related Quantities}

\author{ C.~Bernard,\hskip-0.03in$\,\null^{\rm a}$
\thanks{presented by C.~Bernard at {\it b20:  Twenty Beautiful Years 
of Bottom Physics}, Illinois Institute of Technology, June 29-July 2, 1997}
T.~Blum,\hskip-0.03in$\,\null^{\rm b}$
T.~DeGrand,\hskip-0.03in$\,\null^{\rm c}$
C.~DeTar,\hskip-0.03in$\,\null^{\rm d}$
Steven~Gottlieb,\hskip-0.03in$\,\null^{\rm e}$
U.~M.~Heller,\hskip-0.03in$\,\null^{\rm f}$
J.~Hetrick,\hskip-0.03in$\,\null^{\rm a}$
C.~McNeile,\hskip-0.03in$\,\null^{\rm d}$
K.~Rummukainen,\hskip-0.03in$\,\null^{\rm g}$
R.~Sugar,\hskip-0.03in$\,\null^{\rm h}$
D.~Toussaint,\hskip-0.03in$\,\null^{\rm i}$
and M.~Wingate,\hskip-0.03in$\,\null^{\rm c}$
} 
\address{$^{\rm a}$Department of Physics, Washington University, St.~Louis, MO 63130, USA \\ 
$^{\rm b}$Department of Physics, Brookhaven National Lab, Upton, NY 11973, USA \\ 
$^{\rm c}$Physics Department, University of Colorado, Boulder, CO 80309, USA \\ 
$^{\rm d}$Physics Department, University of Utah, Salt Lake City, UT 84112, USA \\ 
$^{\rm e}$Department of Physics, Indiana University, Bloomington, IN 47405, USA \\ 
$^{\rm f}$SCRI, Florida State University, Tallahassee, FL 32306-4130, USA \\ 
$^{\rm g}$Department of Physics, University of Bielefeld, D-33501 Bielefeld, Germany \\ 
$^{\rm h}$Department of Physics, University of California, Santa Barbara, CA 93106, USA \\
$^{\rm i}$Department of Physics, University of Arizona, Tucson, AZ 85721, USA \\ 
} 

\maketitle

\begin{abstract}
The MILC collaboration computation of heavy-light decay constants
is described. Results for \fB, \fBs, \fD, \fDs and their ratios
are presented.  These results are still preliminary, but the analysis
is close to being completed.  Sources of systematic error,
both within the quenched approximation and from quenching itself,
are estimated, although  the latter estimate
is rather crude. 
A sample of our results is:
$f_B=153\ \pm 10\ {}^{+36}_{-13} \ {}^{+13}_{ -0}\ {\rm MeV}$,
$f_{B_s}/f_B  = 1.10\ \pm 0.02\ {}^{+0.05}_{ -0.03} \ {}^{+0.03}_{ -0.02}$,
and $f_{B}/f_{D_s}  = 0.76\ \pm 0.03\ {}^{+0.07}_{ -0.04} \ {}^{+0.02}_
{ -0.01}$, where the errors are statistical, systematic (within the quenched
approximation), and systematic (of quenching), respectively.
The largest source of error 
comes from the extrapolation to the continuum. 
The second largest source is the chiral extrapolation.  At present,
the central values are based on linear chiral extrapolations; 
a shift to quadratic extrapolations would for example
raise $f_B$ by $\approx\!20$ MeV and thereby make the error within the
quenched approximation more symmetric.
\end{abstract}

\vspace{-8pt}
\section*{Introduction}
\vspace{-5pt}

$B_d$-$\bar B_d$ mixing offers a way to determine the CKM matrix element
$V_{td}$.  Indeed, the mixing parameter $x_d$, defined as
$\Delta M_{B_d}/ \Gamma_{B_d}$,
is given in the Standard Model by
\cite{buras}
\begin{equation}
x_d = \tau_{B_d} {G_F^2 M_W^2\over 6 \pi^2} \eta_B S(x_t) M_{B_d} \xi^2_{B_d}
|V_{td}|^2\ ,
\end{equation}
where $\eta_B S(x_t)$ is perturbatively known, $V_{tb}\cong 1$ is assumed,
and 
\begin{equation}
 \xi_{B_d} \equiv f_{B}\sqrt{B_B}\ ,
\end{equation}
with \fB\ the decay constant of the $B_d$ meson and $B_B$ the corresponding
``bag parameter.''  Since $x_d$ has been measured experimentally, 
an evaluation of the nonperturbative quantity $\xi_{B_d}$ will
determine $V_{td}$.  Similarly, a measurement or bound on the corresponding
mixing parameter $x_s$ for $B_s$ mesons will give information about
$V_{ts}$ if $\xi_{B_s}$ is known, or about $|V_{td}/V_{ts}|$,
given $\xi_{B_s}/\xi_{B_d}$.

The lattice offers a way to compute quantities like $f_B$ and $B_B$
from first principles.  Here, we present a nearly completed
computation by the MILC collaboration of the decay constants 
\fB, \fBs, \fD, \fDs, and their ratios.  
A calculation of the quantities $B_B$ and $B_{B_s}/B_B$,
in the static limit for the heavy quark, is also in progress;
some preliminary results are described in Ref.~\cite{heavyflavor}.

\vspace{-3pt}
\section*{Sources of Error}
\vspace{-5pt}

A key issue in any lattice computation is the reliability of
systematic error estimates.  In order to make these estimates,
we use a range of lattices, both quenched 
(\ie negecting virtual quark loop effects) and
unquenched.
The lattice spacing and volume are varied
over as wide a range as practical.  Lattices used are shown in
Table \ref{lattices}.
We discuss the sources of error in turn.

\begin{aiptable}
\begin{longtable}{lllrrcc}
\caption{Lattices used. The quark mass in virtual loops ($m_{vir}$) is given in
lattice units; where it is absent the lattices are quenched.
$a$ is the lattice spacing, and $\ell$ is the spatial box size.
Lattice Q is used for a finite size check only, and does not 
appear in the rest of the analysis. 
Lattice G was generated by HEMCGC; lattice F, by the Columbia group.}
\label{lattices}
\\\aftercapline
 Name & $\beta$ & $m_{vir}$ & size & \# confs. & $a^{-1}$(GeV) & $\ell$(fm)\\
\afterheadline\endfirsthead
\afterheadline\endhead
\hline\endfoot
\hline\endlastfoot
A& $ 5.7 $&$-$&   $8^3 \times 48$&          200 &$1.3$ &$1.2$\\
B& $ 5.7 $&$-$&   $16^3 \times 48$&        100 &$1.3$ &$2.5$\\
E& $ 5.85 $&$-$&  $12^3 \times 48$&     100 &$1.7$ &$1.4$\\
C& $ 6.0 $&$-$&   $16^3 \times 48$&     100 &$2.0$ &$1.6$\\
Q& $ 6.0 $&$-$&  $ 12^3 \times 48$& 235  &$2.0$ &$1.2$\\
D& $ 6.3 $&$-$&   $24^3 \times 80$&     100 &$3.1$ &$1.6$\\
H& $ 6.52 $&$-$&  $ 32^3 \times 100$& 60  &$4.2$ &$1.5$\\
L& $ 5.445$& $0.025$&  $ 16^3 \times 48$&    100 &$1.3$ &$2.5$\\
N& $ 5.5$& $0.1$&  $ 24^3 \times 64$&    100 &$1.4$ &$3.6$\\
O& $ 5.5$& $0.05$&  $ 24^3 \times 64$&    100 &$1.5$ &$3.2$\\
M& $ 5.5$& $0.025$&  $ 20^3 \times 64$&    100 &$1.5$ &$2.6$\\
P& $ 5.5$& $0.0125$&  $ 20^3 \times 64$&    100 &$1.7$ &$2.4$\\
G& $ 5.6$& $0.01$&   $16^3 \times 32$&    200 &$2.1$ &$1.5$\\
F& $ 5.7$& $0.01$&   $16^3 \times 32$&    49  &$2.4$ &$1.3$\\

\end{longtable}
\end{aiptable}

\vspace{-10pt}
\subsection*{Statistics}
\vspace{-5pt}

The basic objects of interest are the quantum mechanical amplitudes
for the propagation of mesons.  Such amplitudes can be written as a functional 
integral,
which in this case means 
the weighted sum
over all possible paths for the quarks and over all possible configurations
of the gluon field.
Since it is impossible, for practical reasons, to include
all such ``paths,'' one resorts to a statistical sampling, and the answers
therefore have statistical errors. Equivalently, one can say that
we are doing a very large dimensional ($\sim\!10^8$) integral by
Monte Carlo importance-sampling methods.  In practice,
about 100 to 250 gluon field configurations are needed to reduce
the statistical errors to well below the present systematic errors.

\vspace{-8pt}
\subsection*{Isolating the State of Interest}
\vspace{-5pt}

Since we do not know, {\it a priori}, the $B$ meson wave function,
we cannot create a $B$ directly on the lattice.  Instead, we operate
on the vacuum with an ``interpolating field'' which need only have the
same quantum numbers as the $B$.  This creates a superposition
of all states with $B$ quantum numbers: the $B$ itself, radial
excitations, $B+$ glueballs, \etc.  However, on a Euclidean space lattice
states develop in time
as $e^{-Et}$, rather than $e^{-iEt}$.  Therefore, after sufficient
Euclidean time, all the higher energy states will have died away,
leaving a pure $B$ state.

\begin{figure}[b!] 
\vspace{-60pt}
\centerline{\epsfig{file=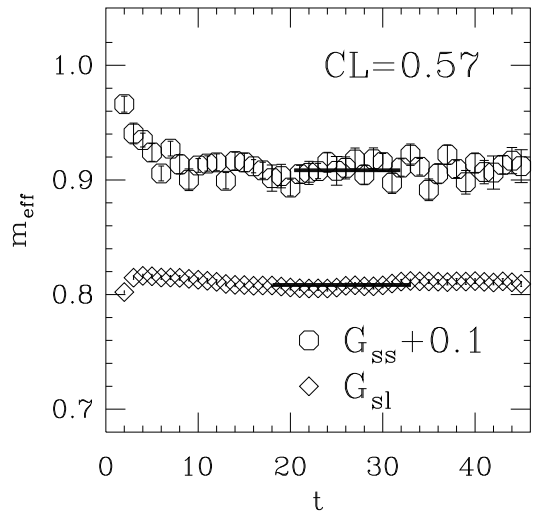,height=3.5in,width=3.5in}}
\vspace{3pt}
\caption{The effective mass $m_{\rm eff}$ {\it vs.}\ Euclidean time for
two meson propagators (on lattice H) created by the same interpolating field.
In $G_{sl}$, the meson is the annihilated by the axial current;
in $G_{ss}$, by a second interpolating field.  For $G_{ss}$, 
$m_{\rm eff}$ has been increased by 0.1 for clarity.  The two
propagators have been fit simultaneously in the ranges shown by the
solid lines. The hopping parameters are $0.1474$ and $0.125$.}
\label{meff}
\end{figure}

The question then becomes: how long a time is ``sufficient?''
A standard approach is to define the ``effective mass,'' $m_{\rm eff}$,
as the instantaneous exponential rate of fall of the $B$ meson propagator.
As the higher energy states die away, the propagator will begin to
fall like a pure exponential, and $m_{\rm eff}$ will approach
the energy (or mass, for zero 3-momentum) of the $B$. Figure~\ref{meff}
shows a typical  plot of $m_{\rm eff}$ {\it vs.}\ time for two propagators.  
(The propagator $G_{sl}$ is needed because one wants to compute
$f_B$, which is proportional to the amplitude for annihilating the
$B$ with an axial current.)
For $t\gtwid 12$, these effective masses show little systematic variation
with time.  However, because of statistical fluctuations, as well as possible
lingering systematic effects, one gets slightly different results
depending on which intervals (with $t\gtwid 12$) are chosen to fit the
propagators.  The variation over the intervals (what we call the
``fitting error'')  is added in quadrature with the naive statistical error
before any further analysis.

\vspace{-8pt}
\subsection*{Quark Mass Extrapolation/Interpolation}
\vspace{-5pt}

$\bullet$ Light-quark Extrapolation.
We compute with light quark masses ($m_q$) in the range
$m_s/3 \ltwid m_q \ltwid 2m_s$.  This is because using physical
$m_u$, $m_d$ would (a) require too much computer time,
(b) require too large a lattice, and (c) introduce spurious
quenching effects
\cite{qchpt}.

The interpolation in $m_q$ to the strange quark mass $m_s$ introduces
little systematic error (assuming one knows how to fix $m_s$ on the lattice).
However, the extrapolation in $m_q$ to physical
$m_{u,d}$ (the chiral extrapolation) is a significant source of error.
That is because one doesn't know {\it a priori} the correct functional
form of the extrapolation. For example, lowest order chiral perturbation theory
predicts that $m^2_\pi$ is a linear function of quark mass.  However,
at the current statistical level, one sees small but
significant deviations from linearity that are not well understood.
In Fig.~{\ref{extrap}(a), the linear fit has very poor confidence level
(taking into account the correlations in the data), despite the fact that
it looks good to the eye.

\begin{figure}[b!] 
\vspace{-56pt}
\centerline{\epsfig{file=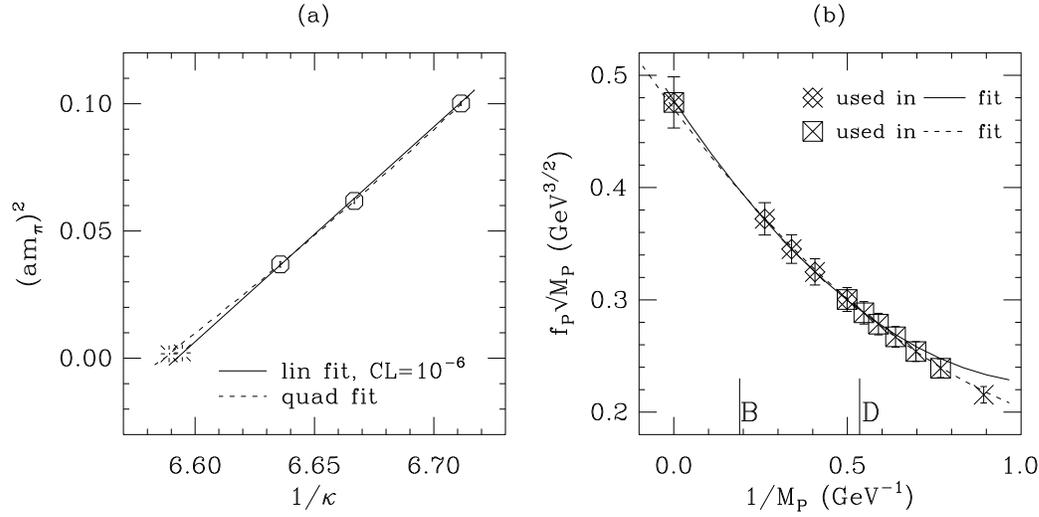,height=3.5in,width=3.5in}}
\vspace{-1pt}
\caption{(a) The lattice $m_\pi^2$ {\it vs.}\ inverse hopping
parameter ( $1/\kappa \approx m_q + {\rm const.}$ ) on lattice D.  
Such fits are used to determine the value of $\kappa$ which corresponds to the physical 
up or down quark mass.
(b) $f_P M_P^{1\over2}$  {\it vs.}\ $1/M_P$ on lattice D, where $M_P$ is the
mass of a heavy-light pseudoscalar meson (with light quark mass 
$\approx m_u, m_d$),
and $f_P$ is its decay constant.  Two different
interpolations to the $B$ and $D$ masses are shown.}
\label{extrap}
\end{figure}

The deviations from linearity
could be due to unphysical effects such as
the finite lattice spacing or the
residual contamination by  radial $B$ excitations.  Even the more
``physical'' cause (chiral logs or higher order 
analytic terms in chiral perturbation theory) are a source of
spurious effects
because quenched chiral logs are in general different from those in the
full theory
\cite{qchpt}.  When we include virtual quark loops, the
theory is at present only ``partially quenched,'' 
since the sea quark mass is not tuned to be equal to the valence mass,
and again spurious chiral logs will be present
\cite{pqchpt}.

For these reasons, we presently
fit quantities like $m^2_\pi$ to their lowest order chiral
form, despite the poor confidence levels.  The systematic error
is estimated by repeating the analysis with quadratic
(constrained) fits, as shown in Fig.~\ref{extrap}(a).
The systematic thus determined is $\le\!10\%$ for decay constants
on all quenched data sets
used to extrapolate to the continuum; usually it is $\ltwid 5\%$.  (After
extrapolation to the continuum, the error is larger: $7\%$ to $15\%$ ---
see below.)
For lattice F, this systematic is very large ($\gtwid 50\%$),
which may be due to the small volume, exacerbated by
virtual quark loop effects.  Lattice F is
dropped from further analysis.

We emphasize that our reasons for choosing linear chiral fits for the
central values are somewhat subjective, and it is possible that we
will switch to quadratic fits in the final version of this work.  
To help us make the choice, we are studying a large sample of new lattices
at $\beta=5.7$, with large volumes up to $24^3$.  On this sample we
have six light quark masses (as opposed to three for each of the lattices
shown in Table \ref{lattices}) and have light-light
mesons with nondegenerate  as
well as degenerate quarks.

$\bullet$ Heavy-quark Interpolation.
Currently practical lattice spacings are in the range
$1\ {\rm GeV} \ltwid a^{-1} \ltwid 4\ {\rm GeV}$.  Thus
$m_B a \gtwid 1$, and it is difficult  to simulate a $B$ meson since its
Compton wavelength is smaller than the lattice spacing.  Our approach
to this problem is to interpolate between heavy-quark masses that
can be better simulated: infinite mass quarks treated
by the static method \cite{eichten} and normal, propagating
quarks with masses lighter than the $b$.  

For the propagating quarks, we correct for the gross lattice artifacts
caused by $m_B a\sim 1$ with some of the techniques of Ref.~\cite{ekm}.
We use the ``EKM norm'' and also shift from pole to kinetic
heavy-quark mass at tadpole-improved tree level.  There are 
further (but numerically less important)  artifacts  which we do not
correct for; we expect them to be eliminated (or at least drastically
reduced) by the continuum extrapolation ($a\to0$) at the end.

One estimate of the systematic error of this approach 
is obtained at fixed lattice spacing by comparing two different 
mass ranges of propagating quarks: ``lighter heavies,'' which give
meson masses in the range 1.25 to 2 GeV, and ``heavier heavies,''
which give
meson masses in the range 2 to 4 GeV.  Figure~\ref{extrap}(b) shows
the interpolations between the static result and the two propagating
mass ranges.  Other estimates of the systematic are available
after one takes the $a\to0$ limit --- see below.

\vspace{-8pt}
\subsection*{Perturbation Theory}
\vspace{-5pt}

The axial current $A_\mu$ with a lattice cutoff is not the same
as the continuum $A_\mu$, but the difference is perturbatively 
calculable, since it comes from physics at the cutoff scale.
At present, we use a (mass independent) one loop matching, with
the scale (``$q^*$'') of the coupling estimated 
along the lines of Ref.~\cite{lepmac}. For propagating Wilson
quarks, the result is, after
tadpole improvement, $q^*=2.32/a$ \cite {bgm}. We estimate
the systematic error by changing $q^*$ by a factor of 2 and reanalyzing.
The error is rather small ($\ltwid 3\%$).

\vspace{-8pt}
\subsection*{Finite Volume}
\vspace{-5pt}

We compare results on lattice A ($\ell=1.2$ fm) with those
on lattice B ($\ell=2.5$ fm).  All other systematics on these two lattices
are the same.  A is smaller than all other lattices used for
our central value; B, much larger.  Therefore the difference
should give a conservative bound on the finite volume error.
This error is $\sim\!3$--$4\%$ on  decay constants and
$\sim\!1$--$2\%$ on  ratios.

\vspace{-8pt}
\subsection*{Extrapolation to the Continuum ($a\to0$)}
\vspace{-5pt}

For any physical quantity Q computed here, we expect (Wilson fermions)
\begin{equation}
Q(a) = Q_{a=0}(1 + a\cM_1 + a^2 \cM_2 + \cdots )\ ,
\end{equation}
but we do not know $\cM_1$ {\it a priori}, nor how it compares
to $\cM_2$.  In practice, we find the slope to be quite large
for the decay constants ($\cM_1 \sim 300$--$650$ MeV), with \fBs\ the
worst offender.  This leads to rather large extrapolation errors
($\sim\!12$--$27\%$).  The ratios of decay constant
are  much better behaved, with
$\cM_1\sim 100$ MeV, and an error of $\sim\!4$--$5\%$.

Figure~\ref{fb}(a) shows several fits of \fB\ {\it vs.}\ $a$ used
to compute the central value and estimate the two largest sources
of systematic error.  The central value is obtained from
a linear fit to the diamonds, which in turn use linear chiral
fits, a lattice scale set by $f_\pi$, and the ``EKM'' corrections
described above. The error of the continuum extrapolation is
estimated by comparing the central value with the result of (1)
a constant fit to the three diamonds with smallest values of $a$,
(2) a linear fit to the squares, which use a different mass shift
(the magnetic --- $m_3$ --- mass minus the pole mass) in the EKM
corrections, and (3) a linear fit to the octagons, which use 
the ``$2\kappa$'' norm --- \ie no EKM corrections.  The 
continuum extrapolation
error is defined as the largest of these three differences.
The difference of the extrapolation
of the crosses (which have a quadratic chiral extrapolation) and the
central value determines the chiral extrapolation error.

\begin{figure}[b!] 
\vspace{-51pt}
\centerline{\epsfig{file=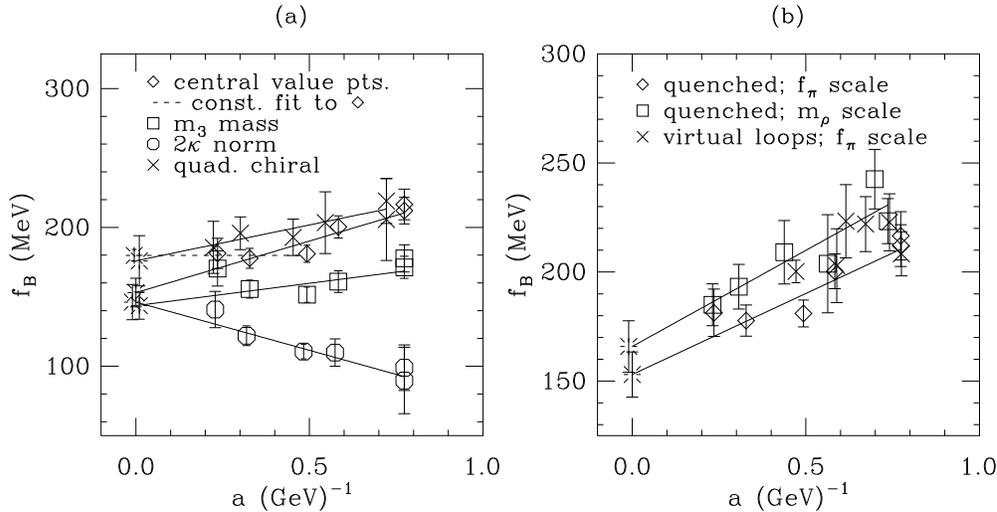,height=3.5in,width=3.5in}}
\vspace{-6pt}
\caption{ (a) Quenched results for \fB\ as a function of lattice spacing.
The linear fit to all the diamonds gives the central value; other
points and fits are used for estimating systematic errors --- see text.
(b) Results for \fB\ as a function of lattice spacing
used for estimating the effects of quenching.  The diamonds
are the same as in (a);  while the squares have the scale fixed by
the mass of the $\rho$ meson.  The crosses (virtual quark loops included)
are not extrapolated to the continuum. 
}
\label{fb}
\end{figure}

\vspace{-8pt}
\subsection*{Effects of Quenching}
\vspace{-5pt}

The quenched approximation has one great advantage: it saves an enormous
amount of computer time.  However, it is not a true approximation,
since there is no perturbative expansion of the full theory
for which the quenched approximation is the first term.  Thus one
should think of it only as a model.  It is a good
model since it
has confinement and chiral symmetry breaking built in,
is completely relativistic,
seems to get the low lying hadronic spectrum right
to $\sim\!5$ to $10\%$, and
can be shown to have small errors in a few cases \cite{cbmglat92}.
But it is a model, nonetheless.  Therefore one must try to estimate 
its errors, and ultimately to move beyond it.

To this end we have repeated our computations on lattices with virtual
quark loops included.  However, we emphasize that such computations
are not yet ``full QCD.''  This is because (1) the virtual quark mass
is fixed and not extrapolated to physical up or down mass (the
theory is partially quenched \cite{pqchpt}), (2) the virtual quark
data is not yet
good enough to extrapolate to $a=0$, and (3) we have two light flavors,
not three.  Thus the virtual quark simulations are used at this point only
for systematic error estimation.

Figure~\ref{fb}(b) shows how the quenching error is estimated.
One estimate is obtained by comparing the
smallest-$a$ virtual quark simulation (lattice G, the cross at
$a=0.47\ ({\rm GeV})^{-1}$) with the quenched simulations, interpolated
to the same value of $a$.  Another estimate can be found by fixing
the lattice scale in the quenched simulations by using $m_\rho$,
instead of $f_\pi$.  In principle, both methods of setting the scale
should be just as good.  However, quenching can affect $m_\rho$
differently from $f_\pi$, so the difference is an estimate
of the systematic.

We emphasize that our quenching error estimate is quite crude at present.
The difference between our virtual quark simulations and full QCD
must be kept in mind.  Further, the comparison of $f_\pi$ and $m_\rho$
scale results tests the quenched approximation only under the assumption that
other systematic errors are well controlled.  The errors
of our continuum extrapolation are large enough to make this last
assumption rather shaky.

\vspace{-4pt}
\section*{Results}
\vspace{-10pt}

Currently, we find:

\begin{eqnarray}
&f_{B_s} = 164\ \pm 9\ {}^{+47}_{ -13} \ {}^{+16}_{ -0}\ \ {\rm MeV}\qquad
&f_{B_s}/f_B = 1.10\ \pm 0.02\ {}^{+0.05}_{ -0.03} \ {}^{+0.03}_{ -0.02_{\phantom{XXX}}}\cr
&f_B = 153\ \pm 10\ {}^{+36}_{-13} \ {}^{+13}_{ -0}\ \ {\rm MeV}\qquad
&f_{D_s}/f_D = 1.09\ \pm 0.02\ {}^{+0.05}_{ -0.01} \ {}^{+0.02}_{ -0.0_{\phantom{XXX}}}\cr
&f_{D_s} = 199\ \pm 8\ {}^{+40}_{ -11} \ {}^{+10}_{ -0}\ \ {\rm MeV}\qquad
&f_{B_s}/f_{D_s} = 0.83\ \pm 0.02\ {}^{+0.06}_{ -0.03} \ {}^{+0.03}_{ -0.00_{\phantom{XXX}}}\cr
&f_D = 186\ \pm 10\ {}^{+27}_{ -18} \ {}^{+9}_{ -0}\ \ {\rm MeV}\qquad
&f_{B}/f_{D_s} = 0.76\ \pm 0.03\ {}^{+0.07}_{ -0.04} \ {}^{+0.02}_{ -0.01}
\end{eqnarray}
\vspace{-0pt}
%
The errors shown are statistical (plus ``fitting''),
systematic (within the quenched
approximation), and systematic (of quenching), respectively.
We note that as experimental measurements of \fDs\ 
improve, the ratios $f_{B}/f_{D_s}$
and $f_{B_s}/f_{D_s}$ may ultimately provide the best way to determine \fB\ and
\fBs.  

The most important issue still under study is whether to switch from
the current linear chiral fits to quadratic ones in determining the
central values. (The difference will still of course be counted
as a systematic error.)  If such a switch were to be made now, it
would raise the central values of \fB, \fBs, \fD\ and \fDs\ by
23, 19, 13, and 14 MeV, respectively.  The systematic
error within the quenched approximation would then become much more symmetric,
with the continuum extrapolation the dominant positive error and the
chiral extrapolation the dominant negative one. 

We thank the Center for Computational Sciences (Oak Ridge), 
Indiana University, 
SDSC (San Diego),
PSC (Pittsburgh), CTC (Cornell) , MHPCC (Maui),  and CHPC (Utah)
for computing resources.
This work was supported in part by the DOE and NSF.

\end{document}